\documentclass{icrc29}
\usepackage{graphicx,amssymb,amsmath,times,url}
\newcommand{\sfrac}[2]{\leavevmode\kern.1em
  \raise.5ex\hbox{\the\scriptfont0 #1}\kern-.1em
  /\kern-.15em\lower.25ex\hbox{\the\scriptfont0 #2}}

\begin{document}
\title[Fitting the HiRes Spectra]{Fitting the HiRes Spectra}
\author[D.R. Bergman, for the HiRes Collaboration] {D.R. Bergman$^a$,
  for the HiRes Collaboration\\
  (a) Rutgers - The State University of New Jersey, Department of
  Physics and Astronomy, Piscataway, New Jersey, USA}
\presenter{Presenter: D.R. Bergman (bergman@physics.rutgers.edu), \ 
  usa-bergman-D-abs3-he14-oral}

\maketitle

\begin{abstract}
  We fit the HiRes ultra-high energy cosmic ray (UHECR) spectrum
  measurements with broken power laws in order to identify features.
  These fits find the previously observed feature known as the Ankle
  at $10^{18.5}$ eV, as well as evidence for a suppression at higher
  energies, above $10^{19.8}$ eV.  We use the integral spectrum and
  the $E_{\sfrac{1}{2}}$ test to identify this high energy suppression
  with the GZK suppression.  Finally, we use a model of uniformly
  distributed extragalactic proton sources together with a
  phenomenological model of the galactic cosmic ray spectrum to
  compare the HiRes spectra to what should be expected from the GZK
  suppression, and to measure how the extragalactic sources must
  evolve and what the input spectral slope must be to fit the HiRes
  data.  Fits using updated spectra will be presented in Pune.
\end{abstract}

\section{Broken Power Law Fits and $E_{\sfrac{1}{2}}$}

\begin{figure}[h]
  \begin{minipage}[t]{0.45\textwidth}
    \includegraphics[width=\textwidth]{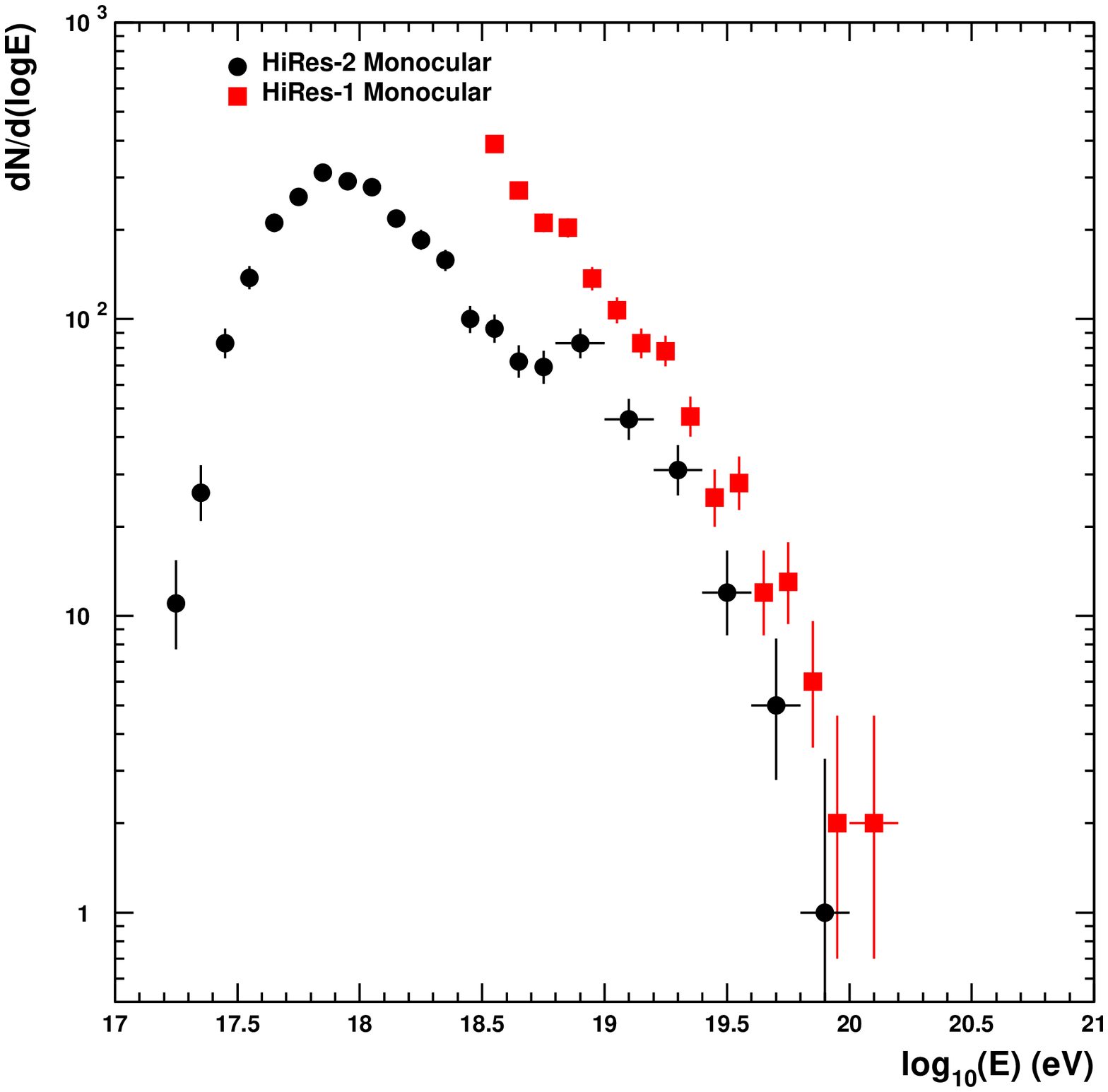}
    \caption{\label{fig:events}  The numbers of events in each bin of
      the HiRes monocular spectrum measurements\cite{ds123}.  The
      HiRes-I measurement includes two empty bins centered at 20.3 and
      20.5.  The HiRes-II measurement includes two empty bins centered
      at 20.1 and 20.3.}
  \end{minipage}\hfill%
  \begin{minipage}[t]{0.45\textwidth}
    \includegraphics[width=\textwidth]{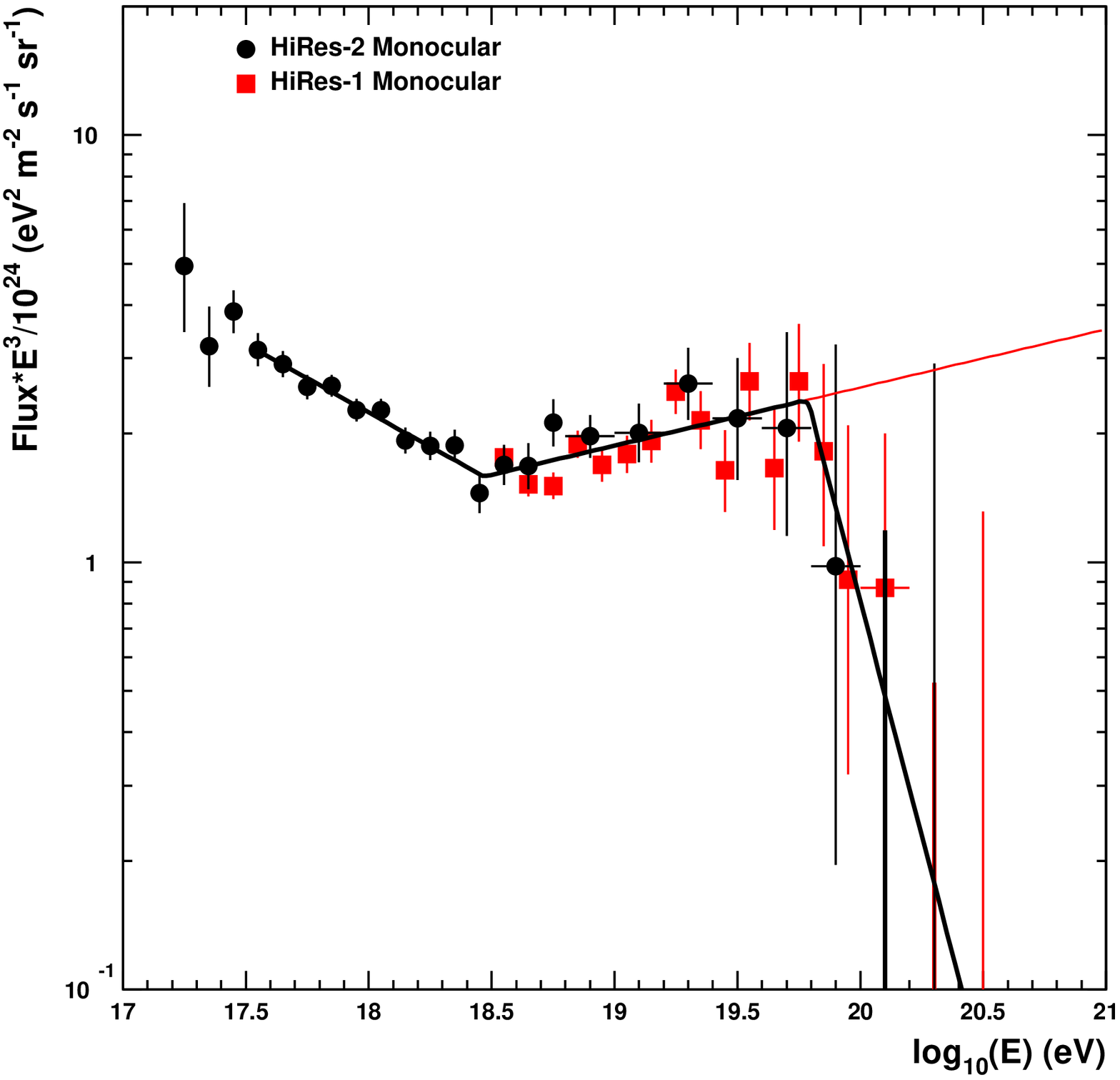}
    \caption{\label{hr12_trilinear_m05} The HiRes spectra
      fit to a BPL with two break points.  The parameters of the fit
      are given in Table~\ref{brokenpowerlawfits}.  The red line is
      used to calculate the significance of the second break point and
      to calculate the expected integral spectrum in the
      $E_{\sfrac{1}{2}}$ calculation.}
  \end{minipage} 
\end{figure}

The HiRes Collaboration has recently released two measurements of the
UHECR flux using monocular observations from its two
sites\cite{ds123}.  Broken power law (BPL) fits made to these
measurements can be used to identify features in the spectrum and to
estimate their statistical significance in the data.  All the fits
presented in this paper are performed using the normalized binned
maximum likelihood method\cite{pdg-maxlik}.  This method requires
comparing the numbers of events expected in a given model to the
number actually observed.  The number of expected events is obtained
from the predicted flux divided by the same exposure used to calculate
the observed flux.  The numbers of events observed by HiRes are shown
in Figure~\ref{fig:events}.  The binned maximum likelihood method also
allows one to use bins in which there were no observed events, but in
which events were expected.  The result of the fits are expressed in
terms of a quality-of-fit parameter $\chi^2$ which approaches a true
$\chi^2$ in the limit of large numbers of events.  The BPL fits are
only made to bins above $10^{17.5}$ eV.  While HiRes-II does have
three bins below this energy, these bins have poor statistics.  We
don't fit these bins in order to avoid biases due to an expected
change in the spectral slope, the Second Knee, at about this energy.
Parameters from fits to the HiRes monocular spectra to a BPL,
$J(E)=CE^{-\gamma}$, with zero, one and two floating break points are
shown in Table~\ref{brokenpowerlawfits}.  The result of the fit in the
two floating break point case is shown in
Figure~\ref{hr12_trilinear_m05}.

\begin{table}
  \caption{\label{brokenpowerlawfits} Parameters found in broken power
      law fits to the HiRes monocular spectra.}
    \begin{center}
      \begin{tabular}{ccccccc}
        \hline
        Fit& $\chi^2$/DOF& $\gamma$& BP& $\gamma$& BP& $\gamma$\\
        \hline
        0 BP&  114/37& 3.12$\pm$0.01&&&&\\
        1 BP& 46.0/35& 3.31$\pm$0.03& 18.45$\pm$0.02& 2.91$\pm$0.03&&\\
        2 BP& 30.1/33& 3.32$\pm$0.04& 18.47$\pm$0.06& 2.86$\pm$0.04&
          19.79$\pm$0.09&5$\pm$1\\ 
        \hline
      \end{tabular}
    \end{center}
\end{table}

The simple power law fit is clearly not a good fit.  Adding one
floating break point gives a much better fit, and the break point
finds the feature known as the Ankle with a very high degree of
statistical significance.  Adding a second floating break point
improves the fit further.  The break point is found to be at
approximately the energy expected of the GZK suppression\cite{GZK}.

The statistical significance of the second break in the spectrum can
be estimated by looking at the reduction in the $\chi^2$ achieved by
adding the break point, or by comparing the number of expected events
above the second break point to what would be expected if the spectrum
continued unabated above the second break point.  The reduction of 16
in the $\chi^2$ for two additional degrees of freedom corresponds to
just under 4$\sigma$ significance in a gaussian fit.  Using the red
line in Figure~\ref{hr12_trilinear_m05}, one would expect to see 28.0
events, where 11 events were actually observed.  The Poisson
probability of observing 11 events or fewer when expecting 28.0 is
$2.4\times10^{-4}$.  For comparison, the area in one tail of a
gaussian distribution outside of 4$\sigma$ (3$\sigma$) is
$3.2\times10^{-5}$ ($1.4\times10^{-3}$).  So the significance of the
high energy suppression is between 3$\sigma$ and 4$\sigma$.

\begin{figure}[h]
  \begin{minipage}[t]{0.45\textwidth}
    \includegraphics[width=\textwidth]{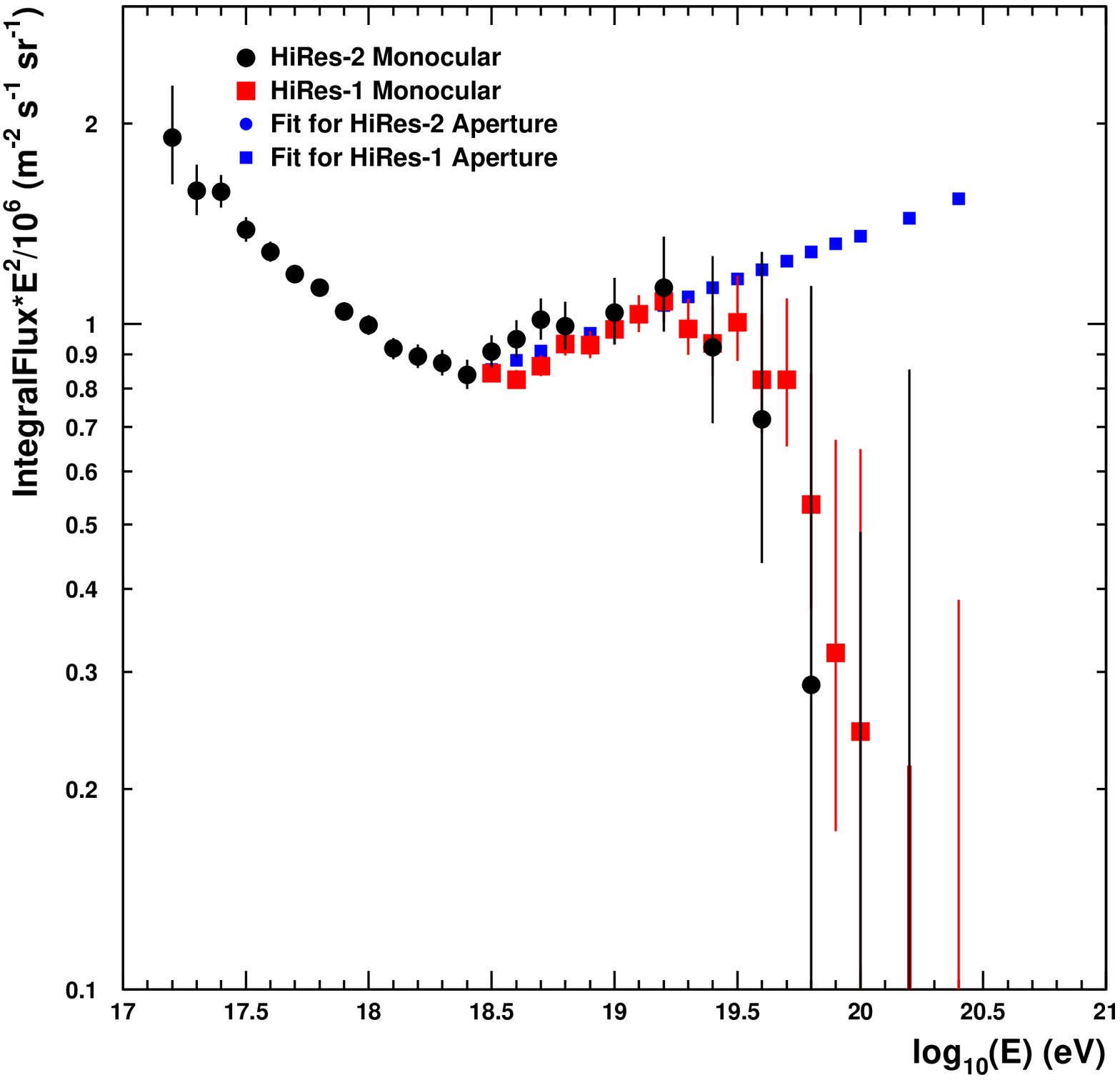}
    \caption{\label{integral-spectrum} The integral spectra derived
      from the differential spectra shown in
      Figure~\ref{hr12_trilinear_m05}.  The blue points represent the
      expected integral spectrum from the red line in that figure.}
  \end{minipage}\hfill%
  \begin{minipage}[t]{0.45\textwidth}
    \includegraphics[width=\textwidth]{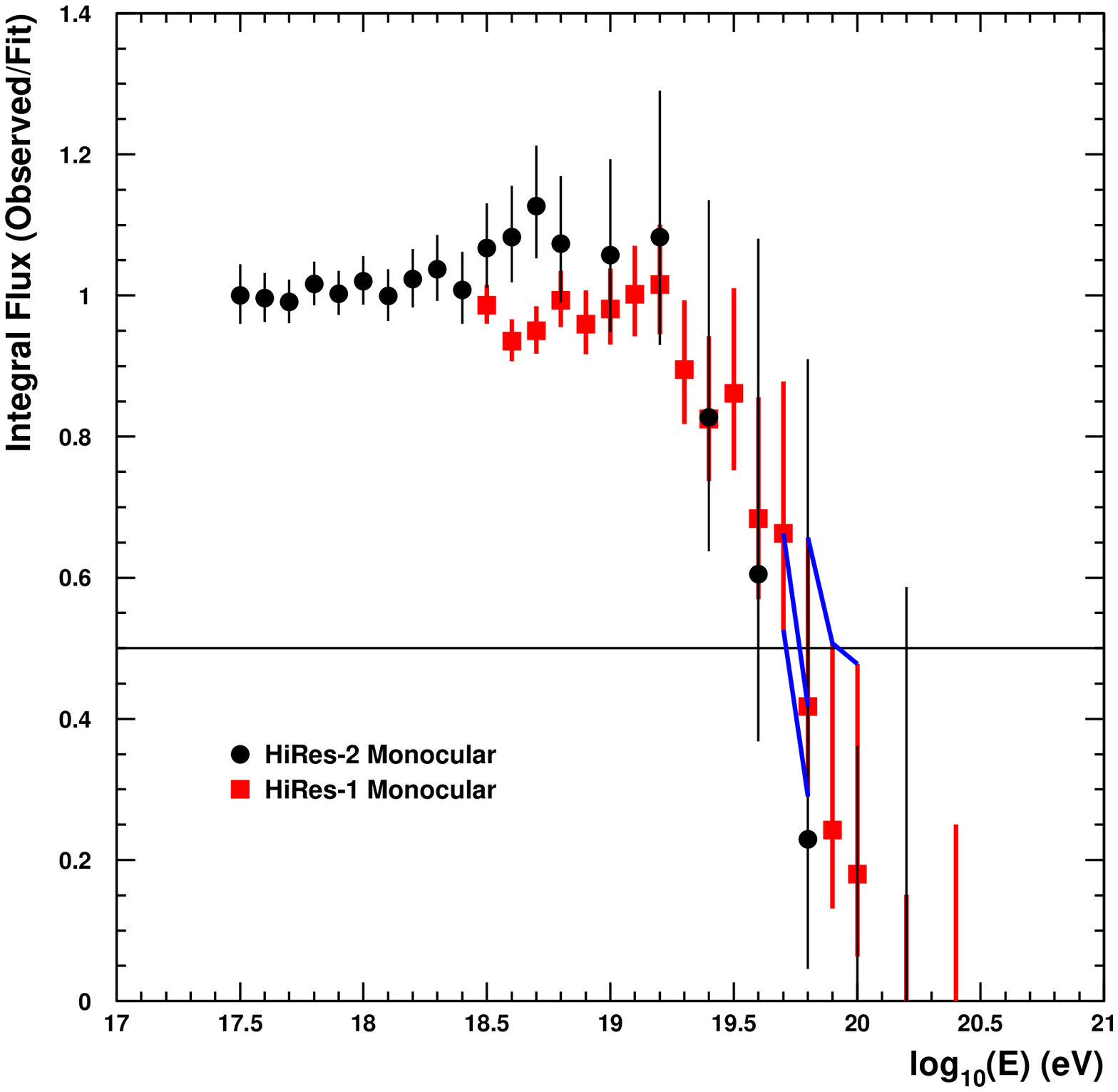}
    \caption{\label{integral-ratio} The observed-to-expected ratio for
      integral spectra.  The value of $E_{\sfrac{1}{2}}$ is obtained
      by a simple interpolation between the HiRes-I points as
      indicated by the blue lines.}
  \end{minipage} 
\end{figure}

Berezinsky {\it et al.}\cite{berezinsky} have suggested measuring the
energy of a break in the UHECR spectrum finding the energy,
$E_{\sfrac{1}{2}}$, at which the observed integral spectrum is half of
what one would expect with no break.  The integral spectrum measured
by HiRes, along with the expected integral spectrum using the red line
in Figure~\ref{hr12_trilinear_m05}, is shown in
Figure~\ref{integral-spectrum}.  The ratio of the observed to the
expected integral spectra is shown in Figure~\ref{integral-ratio}.  By
interpolating between the HiRes-I points in
Figure~\ref{integral-ratio}, we find an experimental value of
$\log_{10}E_{\sfrac{1}{2}}=19.77^{+0.15}_{-0.06}$ ($E$ in eV).
Berezinsky {\it et al.}\cite{berezinsky} have determined
$\log_{10}E_{\sfrac{1}{2}}=19.72$ as what is to be expected for the
GZK suppression for $2.1<\gamma<2.7$.

\section{Uniform Source Model Fits}

Berezinsky {\it et al.}\cite{berezinsky} also calculate the energy
loss rate for UHE protons traveling through the cosmic microwave
background radiation, assuming continuous energy loss through electron
pair production, pion production and universal expansion.  This can be
used to predict the observed spectrum of extragalactic protons for a
given input spectrum.

\begin{figure}[h]
  \begin{minipage}[t]{0.45\textwidth}
    \includegraphics[width=\textwidth]{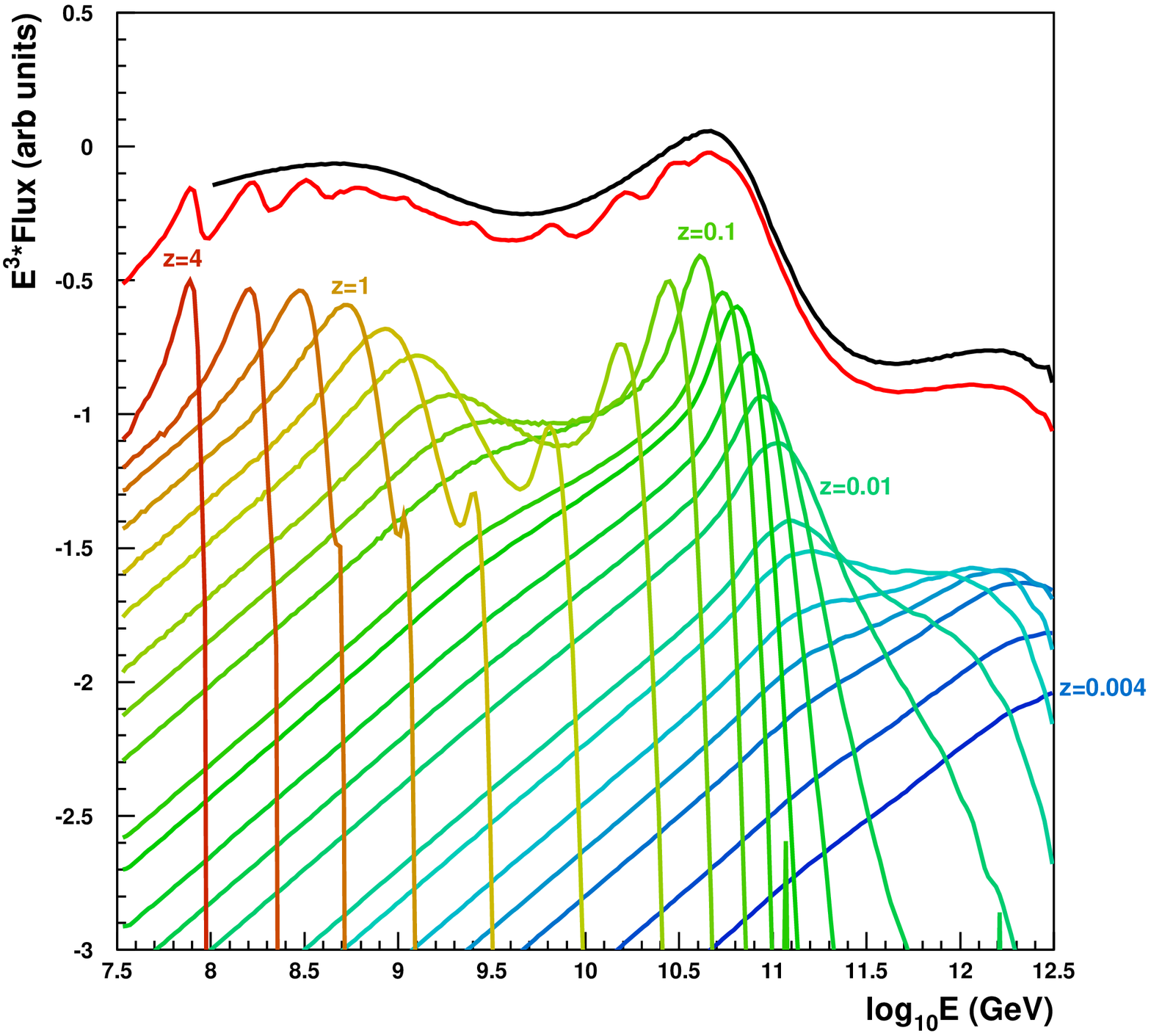}
    \caption{\label{shell-series} Spectrum of extragalactic
      protons from USM with $\gamma=2.4$ and $m=2.5$.  The spectra
      individual shells sum to the red line shown.  A finer series of
      shells sum to the black line.}
  \end{minipage}\hfill%
  \begin{minipage}[t]{0.45\textwidth}
    \includegraphics[width=\textwidth]{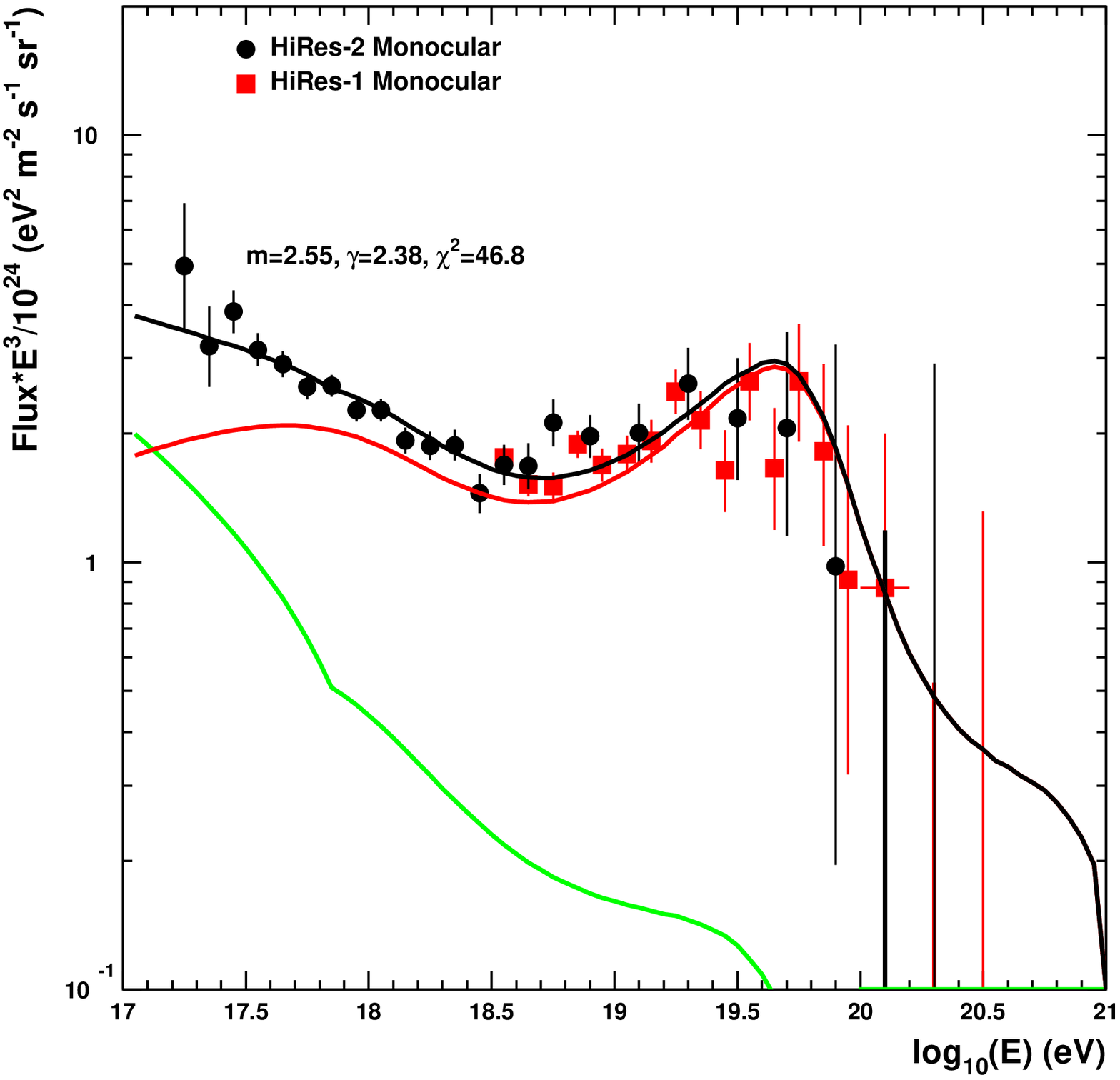}
    \caption{\label{usm-best} Best fit of the USM-plus-Galactic model
      to the HiRes monocular spectra.  The red line shown the
      extragalactic component, the green line shows the galactic
      component and the black line shows the sum.}
  \end{minipage} 
\end{figure}

Since protons with energies above the pion production threshold lose
energy in highly inelastic interactions, we have used the Monte Carlo
method of DeMarco {\it et al.}\cite{DBO} to model this part of the
propagation of extragalactic protons.  Protons from a shell at a given
redshift are propagated from generation to observation (at $z=0$).
Each input energy corresponds to a distribution of observed energies
after such a propagation.  This $E_{\rm in}$--$E_{\rm out}$
distribution convolved with the input spectrum gives the observed
spectrum for protons from sources at a given redshift.  If the
distribution of sources is assumed to be uniform at any given
redshift, but to evolve with redshift as $(1+z)^m$, one can combine
the spectra from different shells as shown in for a coarse,
logarithmic series of shells in Figure~\ref{shell-series}.

Because the HiRes-II spectrum extends to fairly low energies, one
should take into account an additional galactic component when fitting
to this USM model.  We made the simple phenomenological assumption
that the extragalactic and galactic components of the spectrum are
expressed, respectively, in the light (protons) and heavy (iron)
components of a composition measurement.  We use fits to the HiRes
Prototype/MIA\cite{hrmia} and HiRes Stereo\cite{hrstereo} composition
measurements with respect to QGSJet proton and iron expectations to
determine the relative sizes of the extragalactic and galactic
components.  Then, for a given set of parameters for the extragalactic
UHECR spectrum, we can add the appropriate galactic cosmic ray flux.

We varied the input spectral slope, $\gamma$, and evolution parameter,
$m$, to find the best fit of the HiRes monocular data to this
USM-plus-Galactic model.  The best fit spectrum is shown in
Figure~\ref{usm-best}.  The final result for the extragalactic USM
model is $\gamma=2.38\pm0.035{\rm (stat)}\pm0.03{\rm (syst)}$, $m =
2.55\pm0.25{\rm (stat)}\pm0.30{\rm (syst)}$, where the systematinc
uncertainties come from varying the extragalactic/galactic ratio
within the limit allowed by the composition measurements.  

The fit works best in the region on either side of the Ankle, with the
fall at lower energies, into the ankle, primarily determining $m$, and
the rise at higher energies, out of the Ankle, primarily determining
$\gamma$.  The fit does not work as well in the region just below the
GZK suppression.  This could be due to some sources having a maximum
energy below the GZK threshold.  The is also little sign of a Second
Knee at around $10^{17.5}$ eV, though the HiRes-II data has little
statistical power in this region.

\section{Conclusion}

The HiRes detector has observed the Ankle and has evidence for a
suppression at higher energies above $10^{19.8}$ eV.  The energy for
this high energy suppression agrees with what is expected from the GZK
suppression according to the $E_{\sfrac{1}{2}}$ test.  The observed
spectra are well fit by a USM-plus-Galactic model, which finds an
input spectral slope for extragalactic protons of
$\gamma=2.38\pm0.035{\rm (stat)}\pm0.03{\rm (syst)}$, and an evolution
parameter $m = 2.55\pm0.25{\rm (stat)}\pm0.30{\rm (syst)}$.

This work is supported by US NSF grants PHY-9321949, PHY-9322298,
PHY-9904048, PHY-9974537, PHY-0098826, PHY-0140688, PHY-0245428,
PHY-0305516, PHY-0307098, and by the DOE grant FG03-92ER40732. We
gratefully acknowledge the contributions from the technical staffs of
our home institutions. The cooperation of Colonels E.~Fischer and
G.~Harter, the US Army, and the Dugway Proving Ground staff is greatly
appreciated.


\begin{thebibliography}{99}
\bibitem{ds123} R. Abbasi {\it et. al.}, To appear in Phys. Lett. B.,
  astro-ph/0501317; see also
  \url{http://www.physics.rutgers.edu/~dbergman/HiRes-Monocular-Spectra.html}.
\bibitem{pdg-maxlik} S.~Eidleman {\it et al.}, Phys.~Lett. {\bf B592},
  1 (2004).
\bibitem{GZK} K.~Greisen, Phys. Rev. Lett. {\bf 16}, 748 (1966);
  G.T.~Zatsepin and V.A.~K'uzmin, Pis'ma Zh. Eksp. Teor.  Fiz.  {\bf
    4}, 114 (166) [JETP Lett. {\bf 4}, 78 (1966)].
\bibitem{berezinsky} V.~Berezinsky, A.Z.~Gazizov and S.I.~Grigor'eva,
  hep-ph/0204357.
\bibitem{DBO} D.~De Marco, P.~Blasi and A.V.~Olinto, Astropart.~Phys.
  {\bf 20} 53 (2003).  
\bibitem{hrmia} T.~Abu-Zayyad {\it et al.}, Phys. Rev. Lett. {\bf
    84}, 4276, (2000).
\bibitem{hrstereo} R. Abbasi {\it et al.}, Astrophys.~J. {\bf 622},
  910 (2005).
\end{thebibliography}
\end{document}